\documentclass{article}    

\usepackage{graphicx}

\title{\textbf{Timing of the Penrose R-Process}}  
\author{Richard Mould\footnote{Department of Physics and Astronomy, State University of New York, Stony Brook,
\mbox{New York} 11794-3800; http://nuclear.physics.sunysb.edu/ \~{}mould}}  
\date{}    
   
\begin{document}             

\maketitle              

\begin{abstract}

The Penrose reduction theory is applied to a particle/detector interaction where the time that it takes for a
particle capture (and reduction of the original state) is constrained by the requirement that $ \Delta t =  \hbar/\Delta
E$.  The usual interpretation of this equation is found to conflict with the Hamiltonian dynamics of the interaction. 
Another interpretation of this equation seems to solve that problem, but then another difficulty arises. We
conclude that the Penrose theory cannot reflect the timing implicit in the Hamiltonian dynamics unless it can be amended 
to take account of the probability current flow between competing states.  

\end{abstract}

\section*{Introduction}

	Penrose assumes that Nature will not tolerate significant differences in geometry between two competing eigenstates in a
quantum mechanical system.  If the metrical uncertainty between the states is greater than the Planck length, then
according to Penrose, there will be an ``objective" collapse (\textbf{R}-process) of the wave function that selects one
component of the superposition and drives the other to zero \cite{RP}.  This collapse is coupled to the onset of
environmental decoherence between the components, rendering them locally incoherent.  

	The timing of a collapse is said to depend on the difference between the gravitational energy $\Delta E$ of the competing
eigenstates including their environments. The time $\Delta t$ allowed for the \textbf{R}-process to occur is then given
by  $\hbar/\Delta E$.  In the macroscopic case to be considered, the difference in gravitational energy of the competing
environments will dominate the effect, guaranteeing a virtually instantaneous decay of the state function as classically
expected \mbox{(ref.\ 1, pp. 341-342).}

\section*{The Interaction}

In another paper \cite{RM}, the author considers an interaction in which a particle/detector system is given by $\Phi(t)
= exp(-iHt)\psi _iD_i$, where $\psi _i$ and $D_i$ are the initial states of the incoming particle and the detector, and
$H$ is the Hamiltonian divided by $\hbar$. Because $H$ includes an interaction term between $\psi$ and $D$, the resulting
state 
$\Phi(t)$ is an entanglement in which the particle variables and the detector variables are not separable.  However,
$\Phi(t)$ is a superposition (in the representation considered) so we can group its components in any way that we like.  I
form two major components: (1) those detector states $D_0$ in which there is no particle capture, and (2) those detector
states $D_1(t)$ in which there is a particle capture.  Let $\psi(t)$ represent the free particle as a function of time,
including all the incoming and scattering components.   It is therefore entangled with $D_0$.  But since the detector is
macroscopic, we may approximate $D_0$ to be a constant that can be factored out of $\psi(t)$.  The captured particle is
included in the component $D_1(t)$.  This gives

\begin{equation}
\Phi(t \ge t_0) = \psi(t)D_0 + D_1(t)
\end{equation}
where $D_1(t)$ is equal to zero at $t_0$ and increases in time, whereas $\psi(t)D_0$ is normalized to one at $t_0$ and
decreases in time.

When the environments $E_0$ and $E_1$ are attached to these components the $\Delta E$ associated with these macroscopic
components is very large; so, the Penrose theory predicts an ``immediate" stochastic
reduction.  But that cannot be correct, for the Hamiltonian has other plans.  

	The Hamiltonian requires that the superposition in eq.\ 1 will last as long as is necessary for the interaction to run its
course.  The timing of a collapse (i.e., a particle capture) is determined by the probability current flowing from the
first to the second component, and that is determined by the rate at which the particle wave approaches and overlaps the
detector at each moment.  For low-level radioactive emissions yielding particle waves that are spread widely over space,
the interaction in eq.\ 1 might last for hours or days.  In that case the superposition will last for hours or days.  So
the `time' of a particle capture in this treatment is not a function of decoherence or gravitational thresholds.  It is,
\emph{and should be}, a function of the parameters of the incoming particle and the cross section of the particle with the
detector. The dynamics of the Hamiltonian therefore conflict with the predictions of the Penrose theory.

\section*{Alternative Interpretation}

There is another difficulty with the above interpretation of  $\hbar/\Delta E$.  It is said that a reduction occurs
immediately because of the great difference in gravitational energy between the first and second components in eq.\ 1.  But
``immediately" means at time $t_0$, and at time $t_0$ the second component is equal to zero.  How can the second component
be gravitationally competitive with the first component if it is equal to zero?  

Normally the decay time of a quantum mechanical system is given by  $\hbar/\Delta E$, where $\Delta E$ is its Heisenberg
spread in energy.  In the above example $\Delta E$ would then correspond to the spread in the energy of the incoming
particle wave packet, and the time  $\hbar/\Delta E$ would correspond to the time that it takes for the packet to pass
over the detector.  This `alternative' interpretation of $\Delta E$ would give the right result.  The trouble is, the
expression $\Delta t = \hbar/\Delta E$ is a minimum condition, so $\Delta t$ might be much larger than  $\hbar/\Delta E$. 

Imagine that the incoming particle is a photon that is divided into two equal pulses by a half-silvered mirror.  Let both
pulses be directed toward the detector, which is represented in the accompanying figure by a shaded rectangle; and assume
that the total cross section is such that there is a probability of 1.0 that the particle will be captured.  Then the
uncertainty of the time of interaction is given by $\Delta t$ which is approximated in the figure as the time interval
between the pulses.  This interval can be made indefinitely large, so the interaction time can be made indefinitely large
without changing the uncertainty in energy $\Delta E$.

\begin{figure}[h]
\centering
\includegraphics[scale=0.8]{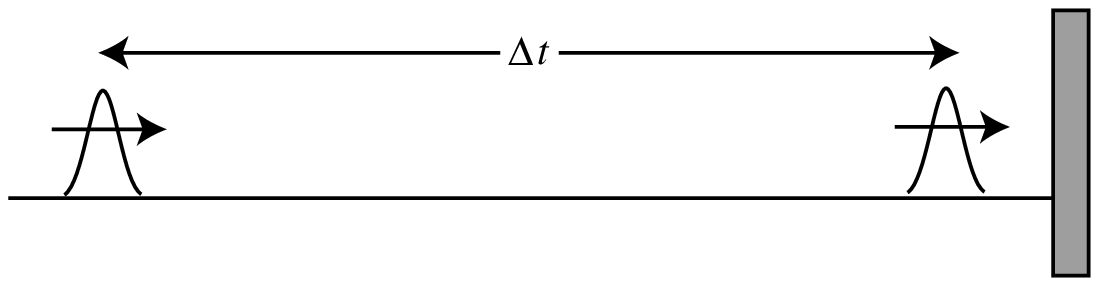}
\end{figure}

If the particle is not captured when the first pulse goes by, it will be captured when the second pulse arrives at the
detector.  Since the theory prescribes that a reduction must occur within a time  $\hbar/\Delta E < \Delta t$, a
reduction will necessarily occur \emph{between} the two pulses according to our second (alternative) interpretation of 
$\hbar/\Delta E$.  This is certainly not correct.  Another way of putting this is to say that the second component in eq.\
1 would be stochastically chosen at a time when there is no probability current flowing into it -- also not correct.  So
the second interpretation of  $\hbar/\Delta E$ is just as problematic as the first. This argument disqualifies any
interpretation of  $\hbar/\Delta E$ because $\Delta t$ in the figure can be made indefinitely large.   

The Penrose theory attributes state reduction (\textbf{R}-process) to a gravitational tension that exists between competing
components in a quantum mechanical superposition.    However, the theory does not work as a function of time
because it cannot generally replicate Hamiltonian dynamics.  If the theory is to be modified in such a way that it does
follow the Hamiltonian, it must find a way to introduce \emph{probability current} as a determining
influence.  The importance of probability current to \emph{any} theory of state reduction is explained in
\mbox{ref.\ 2}.  A Penrose-like gravitational tension must have something to do with the dynamics between states, and
not just their respective static magnitudes.

\end{document}